\documentclass[11pt,a4paper]{article}
\usepackage{graphicx,caption}
\usepackage{ifpdf}
\usepackage{xspace}
\usepackage{jheppub}
\usepackage{dsfont}
\usepackage{bm}
\usepackage{amsfonts}
\usepackage{amsmath}
\usepackage{amssymb}
\usepackage{bbold}
\usepackage{xurl}
\usepackage{hyperref}
\usepackage{axodraw2}
\usepackage{pstricks}
\usepackage{relsize,exscale,scalefnt,anyfontsize,cases}
\usepackage[utf8]{inputenc}
\usepackage{slashed}

\graphicspath{{./figs/}}

\title{Heavy flavor production under a strong magnetic field}

\author[a]{Shile Chen,}
\author[b]{Jiaxing Zhao,}
\author[a]{Pengfei Zhuang}

\affiliation[a]{Department of Physics, Tsinghua University, Beijing 100084, China}
\affiliation[b]{SUBATECH, Universit\'e de Nantes, IMT Atlantique, IN2P3/CNRS, 4 rue Alfred Kastler, 44307 Nantes cedex 3, France}

\emailAdd{csl18@tsinghua.org.cn}
\emailAdd{jzhao@subatech.in2p3.fr}
\emailAdd{zhuangpf@mail.tsinghua.edu.cn}

\abstract{The magnetic field created in high energy nuclear collisions will affect the dynamical processes in the QCD medium, especially the heavy quark production that happens in the initial stage of the collisions. We calculate in a strong magnetic field the heavy quark production cross section for the elementary process $gg\to Q\bar Q$ at leading order and the corresponding transverse momentum distribution in nucleus-nucleus collisions. In comparison to the QED process, the heavy quark production is dominated by the unique QCD channel with gluon self-interaction. Due to the dimension reduction of quark phase space in a strong magnetic field, the production is concentrated in a very narrow energy region above the threshold. Since the translation invariance is broken, the production becomes anisotropic in magnetic field. }

\keywords{heavy quark production; strong magnetic field; Landau levels.}

\begin{document}

\maketitle
\setcounter{footnote}{0}
\renewcommand{\thefootnote}{\arabic{footnote}}

%\newpage

\section{Introduction}

It is widely believed that the strongest electromagnetic field in nature is generated in high energy nuclear collisions~\cite{Kharzeev:2007jp,Skokov:2009qp,Voronyuk:2011jd,Deng:2012pc,Tuchin:2013ie}. The maximum of the magnetic field can reach $5m_{\pi}^2\sim 0.1\ \rm GeV^2$ in Au+Au collisions at top RHIC energy and almost $70m_\pi^2\sim 1\ \rm GeV^2$ in Pb+Pb collisions at LHC energy~\cite{Deng:2012pc,Tuchin:2013ie}, where $m_\pi$ is the pion mass. This external field induces many novel transport phenomena in the evolution of the QCD medium created in the collisions, such as the chiral magnetic effect~\cite{Kharzeev:2007jp,Fukushima:2008xe}, photoproduction of dileptons and quarkonia in peripheral and ultra-peripheral collisions~\cite{Zha:2018ytv,ALICE:2013wjo,ALICE:2015mzu,STAR:2018ldd,STAR:2019wlg}, splitting of $D^0$ and $\bar D^0$ directed flows~\cite{Das:2016cwd,STAR:2019clv,ALICE:2019sgg}, and spin-polarized difference between $\Lambda$ and $\bar \Lambda$~\cite{Muller:2018ibh,Guo:2019joy}. However, it is still difficult to find definite and clean evidence of the magnetic field from the final stage observables. This may lie in the weak signals of the spin-related quantum fluctuations or the short lifetime of the magnetic field~\cite{Deng:2012pc,Tuchin:2013ie,Wang:2021oqq,Yan:2021zjc,Chen:2021nxs}. Different from the light quarks which are produced in the hot medium where the magnetic field is already weak, heavy quarks are produced at the very early stage of the collisions and carry the information of the strongest magnetic field.

Most of the studies on heavy flavors in magnetic field are on their static properties, such as the mass and shape of the open and hidden heavy flavor states~\cite{Marasinghe:2011bt,Alford:2013jva,Cho:2014exa,Iwasaki:2021nrz,Guo:2015nsa}, see for instance a recent review~\cite{Zhao:2020jqu}. The magnetic field will affect also dynamic processes associated with heavy quarks, such as the quarkonium dissociation~\cite{Singh:2017nfa,Hasan:2018kvx,Hu:2022ofv}. Since heavy quarks are produced under a strong magnetic field, their production cross section, especially the angular distribution, might be significantly modified by the field. Because the external magnetic field as a vector breaks down the translation symmetry, the charged particle momentum is not conserved, and the normal mode of the quark field is no longer a plane wave. Therefore, the quark external line, propagator, and energy-momentum conservation at the interaction vertex need to be recalculated by solving the Dirac equation in an external magnetic field. For QED, the dynamic processes such as electron-photon (Compton) scattering and pair creation and annihilation are widely investigated~\cite{Herold:1979zz,Melrose:1983svt,Melrose:1983qan,Thompson:2008pp,Kostenko:2018cgv,Kostenko:2019was,Pal:2018nxm}, see also the review~\cite{Harding:2006qn}. For QCD, the gluon self-interaction leads to a new channel for heavy quark production, see the $s-$channel shown in Fig.~\ref{fig1}. In this work, we calculate the heavy quark production under a strong magnetic field.

The paper is organized as follows. In Section~\ref{sec2}, we introduce the solution of the Dirac equation in a magnetic field and construct the Feynman rules associated with quarks. The connection between the quark propagator obtained here and the Schwinger propagator in the frame of proper time is discussed. In Section ~\ref{sec3} we calculate first the cross section for the elementary production process of heavy quark pairs $gg\to Q\bar Q$ at leading order and then the transverse momentum distribution in nucleus-nucleus collisions at LHC energy, under the approximation of lowest Landau level in a strong magnetic field. In Section \ref{sec4} we go beyond the lowest Landau level and discuss the contribution from the higher Landau levels. We summarize in Section~\ref{sec5}.

%%%%%%%%%%%%%%%%%%%%%%%%%%%%
\section{Dirac equation under a magnetic field}
\label{sec2}
%%%%%%%%%%%%%%%%%%%%%%%%%%%%
Different from thermodynamic functions like potential and energy density which are shown as closed Feynman diagrams controlled only by particle propagators, dynamical processes correspond to open Feynman diagrams with particle external lines and propagators. Since gluons and ghosts do not carry electric charge, their Feynman rules are not changed in an external magnetic field. The Feynman rules associated with quarks are controlled by the Dirac equation
\begin{equation}
\label{Dirac}
\left[i\gamma^\mu \left(\partial_\mu+iqA_\mu\right)-m\right]\psi=0,
\end{equation}
where $m$ is the quark mass, $q$ the quark electric charge, and $A_\mu$ the electromagnetic potential. We consider an external magnetic field $B$ in the direction of $z$-axis and choose the Landau gauge with $A_0=0$ and ${\bm A}=Bx{\bm e}_y$. In this case, the momentum along the $x$-axis is not conserved. Taking into account the Landau energy levels for a fermion moving in an external magnetic field,
\begin{equation}
\label{Landau}
\epsilon^2 = p_z^2+\epsilon_n^2,\ \ \epsilon_n^2 = m^2+p_n^2,\ \ p_n^2=2n|qB|
\end{equation}
controlled by the quantum number $n = 0,1,2,\cdots$, the stationary solution of the Dirac spinor can be written as~\cite{Melrose:1983svt,Kostenko:2018cgv,Kostenko:2019was},
\begin{equation}
\label{spinor}
\psi_{n,\sigma}^-(x,p) = e^{-ip\cdot x} u_{n,\sigma}({\bm x},p),\ \ \ \ \psi_{n,\sigma}^+(x,p) = e^{ip\cdot x} v_{n,\sigma}({\bm x},p)
\end{equation}
with positive-energy spinors
\begin{eqnarray}
\label{positive}
u_{n,-}({\bm x},p) &=& \frac{1}{f_n}\left[\begin{matrix}-ip_z p_n \phi_{n-1}\\(\epsilon+\epsilon_n)(\epsilon_n+m)\phi_n\\-ip_n(\epsilon+\epsilon_n)\phi_{n-1}\\-p_z(\epsilon_n+m)\phi_n\end{matrix}\right],\nonumber\\
u_{n,+}({\bm x},p) &=& \frac{1}{f_n}\left[\begin{matrix}(\epsilon+\epsilon_n)(\epsilon_n+m)\phi_{n-1}\\-ip_z p_n \phi_n\\p_z(\epsilon_n+m)\phi_{n-1}\\ip_n(\epsilon+\epsilon_n)\phi_n\end{matrix}\right]
\end{eqnarray}
and negative-energy spinors
\begin{eqnarray}
\label{negative}
v_{n,+}({\bm x},p) &=& \frac{1}{f_n}\left[\begin{matrix}-p_n(\epsilon+\epsilon_n)\phi_{n-1}\\-ip_z(\epsilon_n+m)\phi_n\\-p_zp_n\phi_{n-1}\\i(\epsilon+\epsilon_n)(\epsilon_n+m)\phi_n\end{matrix}\right],\nonumber\\
v_{n,-}({\bm x},p) &=& \frac{1}{f_n}\left[\begin{matrix}-ip_z(\epsilon_n+m)\phi_{n-1}\\-p_n(\epsilon+\epsilon_n)\phi_n \\-i(\epsilon+\epsilon_n)(\epsilon_n+m)\phi_{n-1}\\p_zp_n\phi_n\end{matrix}\right],
\end{eqnarray}
where $\sigma=\pm 1$ is the eigenvalue of the Pauli matrix and labels different spin states, the four-momentum is defined as $p_\mu = (\epsilon,0,p_y,p_z)$ with $p_y=aqB$ controlled by the center of gyration $a$,
\begin{equation}
f_n = 2\sqrt{\epsilon\epsilon_n(\epsilon_n+m)(\epsilon_n+\epsilon)}
\end{equation}
is the normalization constant, and $\phi_n$ is the harmonic oscillator wave function,
\begin{equation}
\label{harmonic}
\phi_n(x-a) =\sqrt{\sqrt{|qB|\over\pi}{1\over L^2 2^n n!}}H_n\left(\sqrt{|qB|}(x-a)\right)e^{-|qB|(x-a)^2/2}
\end{equation}
with the Hermite polynomial $H_n(x)$ and the normalization length $L$.

With the Dirac spinors (\ref{positive}) and (\ref{negative}), we can rewrite the modified Feynman rules for quarks. In the following we label a gluon with $(\varepsilon_\mu,k)$ and a quark with $(\sigma,p)$, where $\varepsilon_\mu$ is the gluon polarization, $\sigma$ the quark spin, and $k=(\omega,{\bm k})$ the gluon four momentum. The gluon-quark-antiquark vertex can be written as
\begin{eqnarray}
\label{vertex}
&&-ig \int d^4 x \bar\psi_{n,\sigma}^-(x,p)\gamma_\mu t^c A^\mu_c(x,k)\psi_{n',\sigma'}^-(x,p')\nonumber\\
=&&\frac{-ig}{\sqrt{2\omega L^3}}\int d^4 x e^{-i(p'\pm k -p)\cdot x} \bar u_{n,\sigma}({\bm x},p)\gamma_\mu\varepsilon^\mu u_{n',\sigma'}({\bm x},p'),
\end{eqnarray}
where $t_c$ is the color SU(3) generator, $A_\mu^c(x)$ the gluon field, and $g$ the coupling constant. The quark propagator in coordinate space can be constructed as
\begin{eqnarray}
\label{propagator}
G(x'-x) &=& -i\left({\sqrt{|qB|}L\over 2\pi}\right)^2\int dp_zda\sum_{\sigma,n}\Big[\theta(t'-t)u_{n,\sigma}({\bm x}',p)\bar u_{n,\sigma}({\bm x},p)e^{-ip\cdot(x'-x)} \nonumber\\
&&-\theta(t-t')v_{n,\sigma}({\bm x}',p)\bar v_{n,\sigma}({\bm x},p)e^{ip\cdot(x'-x)}\Big],
\label{prop}
\end{eqnarray}
and the energy and momentum conservation corresponding to the heavy quark production process $gg\to Q\bar Q$ is changed to
\begin{eqnarray}
&& \left[2\pi\delta\left(k_z+k'_z-p_z-p'_z\right)\right]^2 \to 2\pi L\delta\left(k_z+k'_z-p_z-p'_z\right), \nonumber\\
&& \left[2\pi\delta\left(k_y+k'_y-(a'-a)/\lambda^2\right)\right]^2 \to 2\pi L\delta\left(k_y+k'_y-(a'-a)/\lambda^2\right), \nonumber\\
&& \left[2\pi\delta\left(\omega+\omega'-\epsilon-\epsilon'\right)\right]^2 \to  2\pi T \delta\left(\omega+\omega'-\epsilon-\epsilon'\right),
\label{delta}
\end{eqnarray}
where $T$ is the normalization time. Since we have chosen the Landau gauge, there is no momentum conservation in the $x-$direction.

Before calculating the cross section, we prove that the quark propagator \eqref{propagator} is equivalent to the famous Schwinger propagator obtained 70 years ago via the proper-time method~\cite{Schwinger:1951nm}. Aiming to see it clearly, we focus on the Lowest Landau Level (LLL) with $n=0$. In this case, substituting the Dirac spinors into \eqref{propagator}, the propagator becomes
\begin{eqnarray}
G(x'-x) &=& \frac{|qB|}{8\pi^2}e^{-\left[(x-x')^2+(y-y')^2+2i(x'+x)(y'-y)\right]|qB|/4}\int dp_z \nonumber\\
&&\times\Bigg[\frac{e^{ip_{z}(z'-z)}}{\epsilon}\theta(t'-t)\left(\begin{matrix}0&0&0&0\\0&(\epsilon+m)&0&p_{z}\\0&0&0&0\\0&-p_{z}&0&-(\epsilon-m)\end{matrix}\right)\nonumber\\
&&-\frac{e^{-ip_{z}(z'-z)}}{\epsilon}\theta(t-t')\left(\begin{matrix}0&0&0&0\\0&(\epsilon-m)&0&p_{z}\\0&0&0&0\\0&-p_{z}&0&-(\epsilon+m)\end{matrix}\right)\Bigg].
\label{eq.prop_lll}
\end{eqnarray}
The Schwinger propagator in the coordinate space can be expressed as~\cite{Chyi:1999fc}
\begin{eqnarray}
G(x'-x)=\Phi(x,x'){\mathcal G}(x-x')
\label{eq.Schwinger}
\end{eqnarray}
with the translation-invariant propagator
\begin{eqnarray}
{\mathcal G}(x-x') &=& -{1\over (4\pi)^2} \int {ds\over s} \left[m+\gamma \cdot \left(qF\coth(qFs)+qF\right)(x-x')/2\right]\nonumber\\
&&\times e^{-i\left(m^2-q\sigma_{\mu\nu}F^{\mu\nu}/2\right)s-{\rm Tr}\ln\left[\sinh(qFs)/(qFs)\right]/2-i\left[qF\coth(qFs)\right](x-x')^2/4}
\end{eqnarray}
and the symmetry breaking phase factor
\begin{equation}
\Phi(x,x') = e^{-iq\int_x^{x'} d\xi \left[A_\mu+{1\over 2}F_{\mu v}(\xi-x')^v\right] },
\end{equation}
where $F$ is the matrix notation of the external field tensor $F_{\mu v}$. This expression includes the contribution from all Landau levels. Moving to the momentum space, the translation-invariant part gives~\cite{Bandyopadhyay:2016fyd,Singh:2017nfa}
\begin{eqnarray}
\mathcal G (p) = ie^{-p_{\perp}^2/|qB|}\sum_{n=0}^\infty (-1)^n\frac{D_n(|qB|,p)}{p_{||}^2-m^2-2n|qB|}
\end{eqnarray}
with
\begin{eqnarray}
D_n(|qB|,p) &=& (\slashed{p}_{||} +m)\left[(1-i\gamma^1\gamma^2)L_n^0(2p_{\perp}^2/|qB|)-(1+i\gamma^1\gamma^2)L_{n-1}^0(2p_{\perp}^2/|qB|)\right]\nonumber\\
&&-4\slashed{p}_{\perp} L_{n-1}^1(2p_{\perp}^2/|qB|),
\end{eqnarray}
where $p_{\perp} =(p_x,p_y)$ and $p_{||} =(\epsilon,p_z)$ are the momentum components perpendicular and parallel to the magnetic field, and $L_n^m(x)$ are the generalized Laguerre polynomials. Under the approximation of LLL with $n=0$, the above propagator is simplified to
\begin{eqnarray}
\mathcal G_0(p) = ie^{-p_{\perp}^2/|qB|}\left(\slashed{p}_{||}+ m \over p_{||}^2-m^2 \right)(1-i\gamma^1\gamma^2),
\end{eqnarray}
and the corresponding propagator
\begin{eqnarray}
G(x'-x) = e^{-i(y'-y)(x'+x)|qB|/2} \int\frac{d^4p}{(2\pi)^4}\mathcal G_0(p) e^{-ip \cdot (x'-x) }
\end{eqnarray}
is exactly the same as \eqref{eq.prop_lll}. As for the 1st Landau level with $n=1$, we checked the equivalence between \eqref{prop} and \eqref{eq.Schwinger}. Due to the tedious expression, we will not show here the details.

%%%%%%%%%%%%%%%%%%%%%%%%%%%%
\section{Heavy quark production under a magnetic field}
\label{sec3}
%%%%%%%%%%%%%%%%%%%%%%%%%%%%
%%%%%%%%%%%%%%%%%%%%%%%%%%%%
\subsection{Cross section for element process $gg\to c\bar c$}
\begin{figure}[!htb]
\centering
	\includegraphics[width=0.7\textwidth]{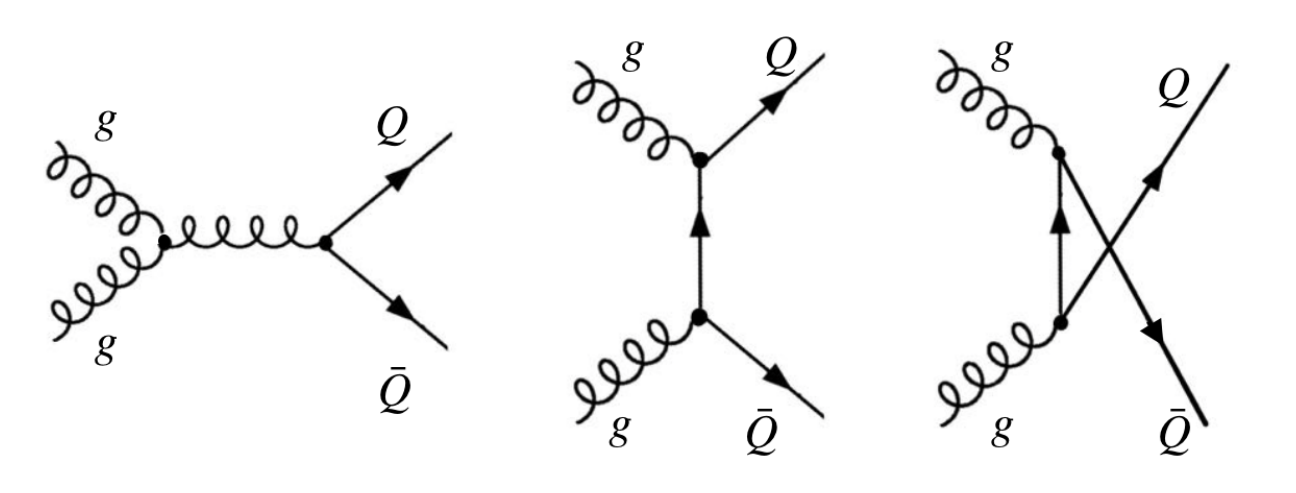}
	\caption{The $s, t$, and $u$ channels of the heavy quark production process $gg\to Q\bar Q$ at the leading order. }
	\label{fig1}
\end{figure}

%%%%%%%%%%%%%%%%%%%%%%%%%%%%
Now we start to calculate the scattering matrix element of heavy quark production in an external magnetic field. At leading order the $s, t$, and $u$ channels of the process $g(k')g(k'')\to Q(p')\bar Q(p'')$ are shown in Fig.~\ref{fig1}. Because the color degree of freedom of gluon is much larger than that of quarks, the contribution from the process $q\bar q\to Q\bar Q$ is much smaller in comparison with $gg\to Q\bar Q$~\cite{Combridge:1978kx,Gluck:1977zm}, we neglect it in this study. We first consider the scattering amplitude for the $s$-channel, which contains a three-gluon vertex and is unique in QCD,
\begin{eqnarray}
S_s &=& {-g^2t_c f^{cab}\over (2\pi)^42\sqrt{\omega'\omega''}L^3} \int d^4 x d^4x' d^4 k{1\over k^2}\bar u_{n',\sigma'}({\bm x},p') \gamma_\mu v_{n'',\sigma''}({\bm x},p'')
\varepsilon'_\rho\varepsilon''_\lambda \nonumber\\
&&\times\left[g^{\mu\rho}(k'+k)^\lambda-g^{\mu\lambda}(k+k'')^\rho+g^{\lambda\rho}(k''-k')^\mu\right]\nonumber\\
&&\times e^{i[(p'+p'')\cdot x-k\cdot(x-x')-(k'+k'')\cdot x']}.
\end{eqnarray}
The integration over $x'$ gives
\begin{eqnarray}
\int d^4x'  e^{ik\cdot x'} e^{-i(k'+k'')\cdot x'}=(2\pi)^4\delta^4 (k-k'-k'').
\end{eqnarray}
Since the cross section is frame independent, we choose the center-of-mass (CM) frame for simplicity. In this case, there is ${\bm k}' + {\bm k}'' = {\bm p}' + {\bm p}'' = 0$, and the amplitude is simplified as
\begin{eqnarray}
S_s &=& -{g^2 t_cf^{cab} \over  \omega L^3} \int d^4 x d^4 k{1\over k^2}\bar u_{n',\sigma'}({\bm x},p') \gamma_\mu v_{n'',\sigma''}({\bm x},p'')
\varepsilon'_\rho\varepsilon''_\lambda e^{-ik\cdot x} \nonumber\\
&&\times \left[g^{\mu\rho}(k'+k)^\lambda-g^{\mu\lambda}(k+k'')^\rho+g^{\lambda\rho}(k''-k')^\mu\right] \delta^3 (k-k'-k''),
\end{eqnarray}
where $\omega$ is the gluon energy in the CM frame. We now do the integration over $k$ and three components $x^0, x^2, x^3$ of $x$. After the integration, the Dirac spinors contain only the component $x^1$, we label it as $x$ in the following expression for simplicity,
\begin{eqnarray}
S_s &=& -{g^2 t_cf^{dab} \over \omega L^3} \int dx {1\over k^2}\bar u_{n',\sigma'}({\bm x},p') \gamma_\mu v_{n'',\sigma''}({\bm x},p'')
\varepsilon'_\rho\varepsilon''_\lambda e^{-ik\cdot x} \nonumber\\
&& \times \left[g^{\mu\rho}(k'+k)^\lambda-g^{\mu\lambda}(k+k'')^\rho+g^{\lambda\rho}(k''-k')^\mu\right] (2\pi)^3\delta^3(E,p_y,p_z),
\label{eq.Ss}
\end{eqnarray}
where the delta function $\delta^3$ means the product of three delta functions for the components $E, p_y$ and $p_z$ shown in \eqref{delta}.
%-----------------------------------------------------
%-----------------------------------------------------

For the $t-$channel, the scattering matrix element can be generally written as
\begin{eqnarray}
S_t &=& ig^2 t^a t^b \left({\sqrt{|qB|}L\over 2\pi}\right)^2{1\over 2\sqrt{\omega'\omega''}L^3}\sum_{n,\sigma}\int d^4 x d^4x' da dp_z
\bar u_{n',\sigma'}({\bm x}',p') \gamma^\nu\varepsilon''_\nu\nonumber\\
&&\times\left[\theta(t'-t)u_{n,\sigma}({\bm x}',p)\bar u_{n,\sigma}({\bm x},p)e^{-ip\cdot(x'-x)}-\theta(t-t')v_{n,\sigma}({\bm x}',p)\bar v_{n,\sigma}({\bm x},p)e^{ip\cdot(x'-x)}\right]\nonumber\\
&&\times\gamma^\mu\varepsilon'_\mu v_{n'',\sigma}({\bm x},p'')e^{-i(k''\cdot x'+k'\cdot x-p''\cdot x-p'\cdot x')}.
\end{eqnarray}
After the time integration,
\begin{eqnarray}
&&\int d t d t'  \theta(t'-t)e^{i[(\epsilon'-\omega'')t'+(\epsilon''-\omega')t-\epsilon(t'-t)]}= 2\pi{\delta(\epsilon'+\epsilon''-\omega'-\omega'')\over \omega''+\epsilon-\epsilon'},\nonumber\\
&&-\int d t d t'  \theta(t-t')e^{i[(\epsilon'-\omega'')t'+(\epsilon''-\omega')t+\epsilon(t'-t)]}= 2\pi{\delta(\epsilon'+\epsilon''-\omega'-\omega'')\over \omega''-\epsilon-\epsilon'},
\end{eqnarray}
the matrix element becomes
\begin{equation}
S_t  = -{ig^2 t^a t^b|qB|\over 4\pi L\sqrt{\omega'\omega''}}\delta(\epsilon'+\epsilon''-\omega'-\omega'')\int da d p_z  \left({I_1I_2\over \omega''+\epsilon-\epsilon'}+{I_3I_4\over \omega''-\epsilon-\epsilon'}\right)
\label{eq.St}
\end{equation}
with
\begin{eqnarray}
I_1 &=& \int d^3 {\bm x}'  \bar u_{n',\sigma'}({\bm x}',p')\gamma^\nu \varepsilon''_\nu u_{n,\sigma}({\bm x}',p)e^{-i({\bm p}'-{\bm k}''-{\bm p})\cdot{\bm x}'},\nonumber\\
I_2 &=& \int d^3 {\bm x}   \bar u_{n,\sigma}({\bm x},p)\gamma^\nu \varepsilon'_\nu v_{n'',\sigma''}({\bm x},p'')e^{-i({\bm p}-{\bm k}'+{\bm p}'')\cdot{\bm x}},\nonumber\\
I_3 &=& \int d^3 {\bm x}'  \bar u_{n',\sigma'}({\bm x}',p')\gamma^\nu \varepsilon''_\nu v_{n,\sigma}({\bm x}',p)e^{-i({\bm p}'+{\bm k}''-{\bm p})\cdot{\bm x}},\nonumber\\
I_4 &=& \int d^3 {\bm x}  \bar v_{n,\sigma}({\bm x},p)\gamma^\nu \varepsilon'_\nu v_{n'',\sigma''}({\bm x},p'')e^{-i({\bm p}''-{\bm p}-{\bm k}')\cdot{\bm x}}.
\end{eqnarray}
The scattering amplitude for the $u-$channel can be obtained from the $t-$channel by exchanging the two initial gluons.

With the scattering amplitudes, the cross section can be calculated directly,
\begin{eqnarray}
\sigma &=& {L^3\over v_{\rm rel}T} \int {L^2dp'_ydp'_z\over (2\pi)^2} {L^2dp''_ydp''_z \over (2\pi)^2}\sum_{\sigma',\sigma''}|S_s+S_t+S_u|^2\nonumber\\
&=& {L^5\over v_{\rm rel}}\left({L\sqrt{|qB|}\over 2\pi}\right)^4\int da' da'' dp'_z dp''_z\sum_{\sigma',\sigma''}|\mathcal M_s+\mathcal M_t+\mathcal M_u|^2(2\pi)^3\delta^3(k'+k''- p'-p'')\nonumber\\
&=& {L^{10}\over 16\pi}{\sqrt s|qB|\over p'_z}\sum_{\sigma',\sigma''}|\mathcal M_s+\mathcal M_t+\mathcal M_u|^2,
\label{eq.crosssection}
\end{eqnarray}
where $v_{\rm rel}=k^{'\mu} k''_\mu/(\omega'\omega'')$ is the relative velocity between the two initial gluons which is reduced to $v_{\rm rel}=2$ in the CM frame, and $\mathcal{M}$ is the scattering matrix element.

Considering the Landau levels induced quark energy difference $\Delta\epsilon^2=2|qB|$ between the ground and first excited state, an often used method in treating the sum over Landau energy levels in a strong magnetic field is the approximation of Lowest Landau Level (LLL) by taking into account only the ground state with quantum number $n=0$~\cite{Hattori:2015aki,Singh:2017nfa}.
In this case, the quark spin is fixed with $\sigma=-1$ for quarks and $\sigma=+1$ for antiquarks, and the Dirac spinors (\ref{positive}) and (\ref{negative}) are simplified as
\begin{eqnarray}
u_{0,-}({\bm x},p) &=& \frac{1}{f_0}\left[\begin{matrix}0 \\ 2m(\epsilon+m)\phi_0(x-a)\\0\\-2mp_z\phi_0(x-a)\end{matrix}\right],\nonumber\\
v_{0,+}({\bm x},p) &=& \frac{1}{f_0}\left[\begin{matrix}0\\-i2mp_z\phi_0(x-a)\\0\\i2m(\epsilon+m)\phi_0(x-a)\end{matrix}\right],
\label{uv}
\end{eqnarray}
and the contraction of the gluon polarization with $\gamma$ matrix becomes
\begin{equation}
\label{contraction}
\gamma^0 \gamma^\mu \varepsilon_\mu = -\left(\begin{matrix}
0 & 0 & \varepsilon_z & \varepsilon_-\\
0 & 0 & \varepsilon_+& -\varepsilon_z\\
\varepsilon_z & \varepsilon_- & 0 & 0\\
\varepsilon_+ & - \varepsilon_z & 0 & 0
\end{matrix}\right).
\end{equation}
Substituting \eqref{uv} and \eqref{contraction} into the scattering amplitude \eqref{eq.Ss}, we obtain
\begin{equation}
S_s = -g^2 t_c f^{cab} {2(2\pi)^3\over L^5}{(\sqrt s/2+m)^2-p'^2_z\over s^2(\sqrt s/2+m)}k'_z \varepsilon'\cdot\varepsilon''\delta^3(E,p_y,p_z),
\end{equation}
with the center-of-mass energy $\sqrt s=\omega$.

Taking the approximation of LLL, the scattering amplitude for the $t-$channel in the CM frame is
\begin{equation}
S_t=-{2(2\pi)^3mg^2 t^a t^b\varepsilon'_z\varepsilon''_z\over s L^5} {-(k''_z-2p'_z) \over s/4+k'_z(k''_z-2p'_z)}e^{-k'^2_\perp/(2|qB|)}\delta^3(E,p_y,p_z)
\end{equation}
with $p_z=p'_z-k''_z$. The scattering amplitude for the $u-$channel can be obtained from the $t-$channel by exchanging the two initial gluons,
\begin{equation}
S_u=-{2(2\pi)^3mg^2 t^a t^b\varepsilon'_z\varepsilon''_z\over s L^5} {(k'_z-2p'_z)\over s/4-k'_z(k'_z-2p'_z)}e^{-k'^2_\perp/(2|qB|)}\delta^3(E,p_y,p_z)
\end{equation}
with $p_z=p'_z-k'_z$.

For the color coefficients, one has $t^at^a=C_2(N_c)I$, $t^at^bt^a = [C_2(N_c)-N_c/2]t^b$ and $[t^a,t^b]=it^c f^{abc}$ with $C_2(N_c)=(N_c^2-1)/(2N_c)$. If the color factors are neglected, the above calculations for the $t-$ and $u-$channel go back to the QED results~\cite{Kostenko:2018cgv}.

The cross section for the $s-$channel can be calculated via \eqref{eq.crosssection},
\begin{eqnarray}
\sigma_{ss}={\sqrt{s}|qB|\over 16\pi}{1\over p'_z}|\mathcal M _s|^2
=C_{ss} {g^4 m^2|qB|k'^2_z |\varepsilon'\cdot\varepsilon''|^2\over \pi s^{7/2}|p'_z|}
\end{eqnarray}
with the final state quark momentum $p'_z=\pm\sqrt{s/4-m^2}$ in the CM frame. Taking into account the color factor $C_{ss}={\rm Tr}([t^a,t^b][t^b,t^a]) = 3/16$, one has
\begin{equation}
\sigma_{ss} = {3g^4 m^2|qB|k'^2_z |\varepsilon'\cdot\varepsilon''|^2\over 16\pi s^{7/2}|p'_z|}.
\end{equation}

Similarly, we can obtain the cross sections
\begin{equation}
\sigma_{tt} = {g^4 m^2|qB||\varepsilon'_z\varepsilon''_z|^2\over 48\pi s^{3/2}|p'_z|}\left|{k'_z+2p'_z\over (p'_z+k'_z)^2+m^2}\right|^2 e^{-k'^2_\perp/|qB|}
\end{equation}
for the $t-$hannel with the color factor $C_{tt} = {\rm Tr}(t^at^bt^bt^a) = 1/12$,
\begin{equation}
\sigma_{uu} = {g^4 m^2|qB||\varepsilon'_z\varepsilon''_z|^2\over 48\pi s^{3/2}|p'_z|}\left|{-k'_z+2p'_z\over (p'_z-k'_z)^2+m^2}\right|^2 e^{-k'^2_\perp/|qB|}
\end{equation}
for the $u-$hannel with the color factor $C_{uu} = C_{tt}$, and
\begin{eqnarray}
\sigma_{ut+tu} &=& -{g^4 m^2|qB||\varepsilon'_z\varepsilon''_z|^2\over 48\pi (s/4)^{3/2}|p'_z|}{4s-4m^2-k'^2_z\over (s/4-k'^2_z)^2+4m^2k'^2_z}e^{-k'^2_\perp/|qB|},\\
\sigma_{st+su} &=& -{3g^4 m^2|qB|k'^2_z[(\varepsilon'_x\varepsilon''_x+\varepsilon'_y\varepsilon''_y)(\varepsilon'^*_ z\varepsilon''^*_z)+|\varepsilon'_z\varepsilon''_z|^2]\over 64\pi\lambda^2 (s/4)^2|p'_z|(\sqrt s/2+m)}\nonumber\\
&&\times {4m^2+k'^2_z-3/4 s\over (s/4-k'^2_z)^2+4m^2k'^2_z}e^{-k'^2_\perp/(2|qB|)},\nonumber\\
\sigma_{ts+us} &=& -{3g^4 m^2|qB|k'^2_z[(\varepsilon'^*_x\varepsilon''^*_x+\varepsilon'^*_y\varepsilon''^*_y)(\varepsilon'_ z\varepsilon''_z)
+|\varepsilon'_z\varepsilon''_z|^2]\over 64\pi (s/4)^2|p'_z|(\sqrt s/2+m)}\nonumber\\
&&\times {4m^2+k'^2_z-3/4 s\over (s/4-k'^2_z)^2+4m^2k'^2_z}e^{-k'^2_\perp/(2|qB|)}\nonumber
\end{eqnarray}
for the mixed processes with the color factors $C_{ut}=C_{tu}={\rm Tr}(t^at^bt^at^b) = -1/96$, $C_{st}=C_{ts}={\rm Tr}([t^a,t^b]t^bt^a)=3/32$ and $C_{su}=C_{us} = -3/32$, where $\varepsilon^*$ is the conjugate of the gluon polarization.

We now consider the average over the initial gluon polarization states. For a gluon propagating in the direction $\hat{\bm k}'=(\sin \theta \cos\phi, \sin \theta \sin\phi,\cos\theta)$ in the CM frame with the magnetic field along the $z-$axis ${\bm B}=B\hat{\bm e}_z$, one can define two independent polarization modes. One is $\varepsilon'_o=(-\cos\theta \cos\phi, -\cos\theta\sin\phi ,\sin \theta)$ which is parallel to the vector $\hat {\bm k}'\times(\hat {\bm k}'\times \hat {\bm B})$, and the other is $\varepsilon'_e=(\sin\phi, -\cos\phi,0)$ which is parallel to the vector $\hat {\bm k}'\times\hat{\bm B}$. For the other initial gluon moving in the opposite direction  $\hat {\bm k}''=-{\bm k}'$, the two polarization modes are $\varepsilon''_o=(\cos\theta \cos\phi, \cos\theta\sin\phi ,-\sin \theta)$ and $\varepsilon''_e=(\sin\phi, -\cos\phi,0)$. With the two polarization modes, the average over the initial state can be expressed as
\begin{eqnarray}
&&\frac{1}{4}\sum_{\rm gluon\ polarization} |\varepsilon'_z\varepsilon''_z|^2 = \frac{\sin^4\theta}{4},\nonumber\\
&&\frac{1}{4}\sum_{\rm gluon\ polarization} |\varepsilon'\cdot\varepsilon''|^2 = \frac{1}{2},\nonumber\\
&&\frac{1}{4}\sum_{\rm gluon\ polarization} (\varepsilon^{'*}_x\varepsilon^{''*}_x+\varepsilon^{'*}_y\varepsilon^{''*}_y)(\varepsilon'_z\varepsilon''_z)+|\varepsilon'_z\varepsilon''_z|^2 = \frac{\sin^2\theta}{4}.
\end{eqnarray}
Under the approximation of LLL, the spin of the final state quarks is fixed, and the sum over the final state polarizations is trivial.

Taking into account all the channels and doing average over the initial state polarization and sum over the final state spin, the total cross section for heavy quark production at leading order is written as
\begin{eqnarray}
\label{sigma}
\sigma(s,B,\theta) &=& {\pi m^2\alpha^2_s|qB|\over s^3\chi}\Bigg\{{3\over 2}\cos^2\theta\left[{1\over2} - {\sin^2\theta\over 1+\sqrt{4m^2/s}}{1+\cos^2\theta-4\chi^2\over \sin^4\theta+16m^2/s\cos^2\theta} e^{-{s\sin^2\theta\over 8|qB|}}\right]\nonumber\\
&&+{2\over 3}\sin^4\theta\bigg [\left({\cos\theta+2\chi\over \left(\chi+\cos\theta\right)^2+4m^2/s}\right)^2
+ \left({-\cos\theta+2\chi\over \left(\chi-\cos\theta\right)^2+4m^2/s}\right)^2, \nonumber\\
&&-{1\over 4}{4\chi^2-\cos^2\theta\over \sin^4\theta+16m^2/s\cos^2\theta}\bigg]e^{-{s\sin^2\theta\over 4|qB|}}\Bigg\},
\end{eqnarray}
where we have used the coupling constant $\alpha_s$ to replace $g$, and $\chi=\sqrt{1-4m^2/s}$ goes to zero at the production threshold $\sqrt s=2m$. The 5 terms here are respectively the contributions from the $s$ channel, $s-$related mixed processes, $t$ channel, $u$ channel, and the $t-$ and $u-$ related mixed processes. For the sake of comparison, we list here also the heavy quark production cross section at the leading order in vacuum~\cite{Combridge:1978kx,Gluck:1977zm},
 \begin{equation}
 \label{vacuum}
\sigma(s)={\pi\alpha_s^2\over 3s}\left[\left(1+{4m^2\over s}+{m^4\over s^2}\right)\log\left (1+\chi \over 1-\chi \right)-\left({7\over 4}+{31m^2\over 4s}\right)\chi \right].
\end{equation}

Before we numerically calculate the total cross section (\ref{sigma}), we analytically discuss its dependence on the polarization angle $\theta$, invariant mass $\sqrt s$ and magnetic field $|qB|$.

1) The cross section depends on not only the magnetic field but also the gluon polarization angle, the magnetic effect is different in different polarization directions. When the polarization is parallel to the magnetic field with $\theta=0, \pi$, there are $\sin\theta=0$ and $\cos\theta=\pm 1$, the contribution from the QED-like processes, namely the $t$ and $u$ channels, disappears, and the production is purely from the $s$ channel which is a unique QCD process due to the gluon self-interaction,
\begin{equation}
\label{0}
\sigma(s,B,0) = \sigma(s,B,\pi) = {3\pi m^2\alpha^2_s|qB|\over 4s^3\chi}.
\end{equation}
When on the other hand the polarization is perpendicular to the magnetic field with $\theta=\pi/2$, there are $\cos\theta=0$ and $\sin\theta=1$, the contribution from the $s$ channel disappears, and the cross section is controlled by the QED-like processes,
\begin{equation}
\label{pi2}
\sigma(s,B,\pi/2) = {14\pi m^2\alpha^2_s|qB|\chi\over 3s^3}e^{-{s\over 4|qB|}}.
\end{equation}
For a general polarization, both the QCD and QED processes contribute to the heavy flavor production.

2) The energy ($\sqrt s$) dependence of the cross section (\ref{sigma}) under a magnetic field looks very different from that in vacuum \eqref{vacuum}. An important reason is the change in the size of the phase space. The used LLL in a strong magnetic field reduces the dimension of the phase space from three to two. This dimension reduction changes the momentum integration element from $\int d^3{\bm p}/p^2$ to $\int d^2{\bm p}/p^2$ which becomes infrared divergent at the threshold energy $\sqrt s=2m$, shown as the common factor $1/\chi$ in \eqref{sigma}. This divergence vanishes in vacuum. A direct consequence of this infrared divergence will be the low-momentum enhancement and high-momentum suppression of heavy-flavor hadrons in high energy nuclear collisions where the created magnetic field is expected to be the strongest in nature. Due to the exponential function of $s$ which is introduced by the internal quark propagator in the magnetic field, the $\theta-$integrated cross section decays very fast with increasing $\sqrt s$.

3) Considering the linear dependence on the magnetic field in the common factor and the exponential function of $1/|qB|$ for the $t$ and $u$ channels, the cross section increases with magnetic field at any polarization angle. It is necessary to point out that the cross section \eqref{sigma} is derived under the assumption of LLL for a strong magnetic field. For a weak magnetic field, one should consider all the Landau energy levels. Therefore, one cannot expect to directly go from \eqref{sigma} to \eqref{vacuum} by taking $B\to 0$.
%------------------------------
\begin{figure}[!htb]
	\centering
	$$\includegraphics[width=0.55\textwidth]{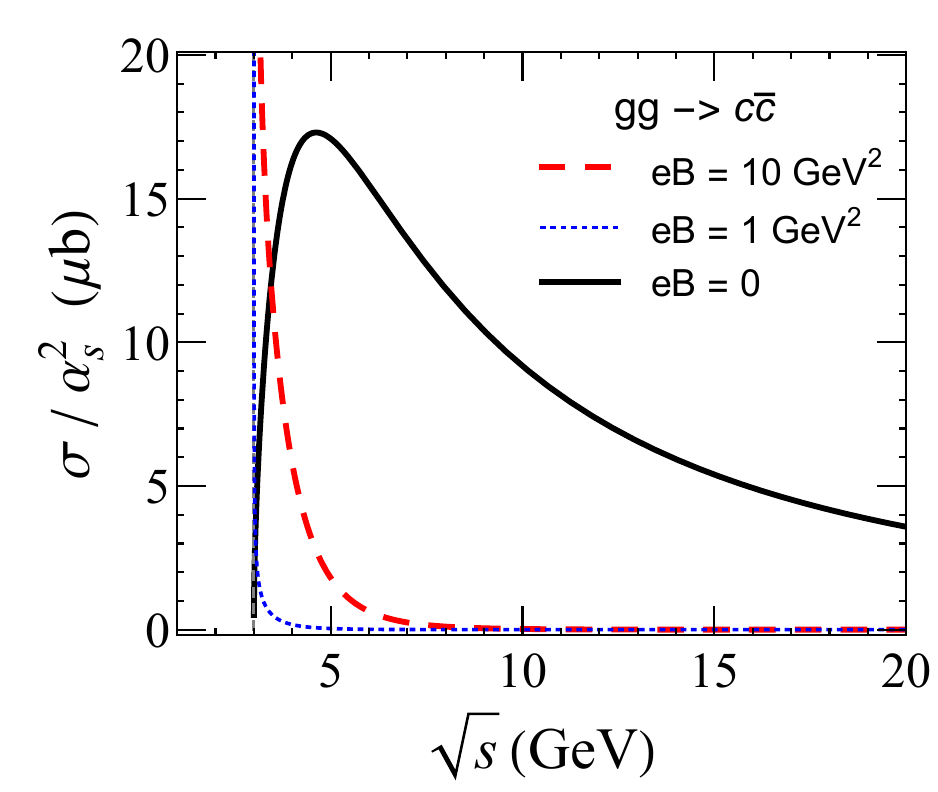}$$
	\caption{The scaled cross section as a function of invariant mass $\sqrt s$ at magnetic field $eB=0$ (solid line), $1$ (dotted line) and $10$ (dashed line) GeV$^2$. }
	\label{fig2}
\end{figure}
%------------------------------

To avoid the coupling constant dependence, we numerically calculate the scaled cross section $\sigma/\alpha_s^2$. We set the charm quark mass $m=1.5$ GeV. The polarization integrated cross section as a function of invariant mass is shown in Fig.~\ref{fig2}. The solid line is the cross section \eqref{vacuum} in vacuum. When the magnetic field is turned on, see the dotted and dashed lines with $eB=1$ and $10$ GeV$^2$, the shape of the integrated cross section is dramatically changed. As we analyzed above, it is divergent at the threshold energy $\sqrt s=2m=3$ GeV and then drops down with the invariant mass very fast. When the strength of the magnetic field increases, the cross section expands but still concentrates in a narrow region above the threshold.

\begin{figure}[!htb]
	\centering
	\includegraphics[width=0.9\textwidth]{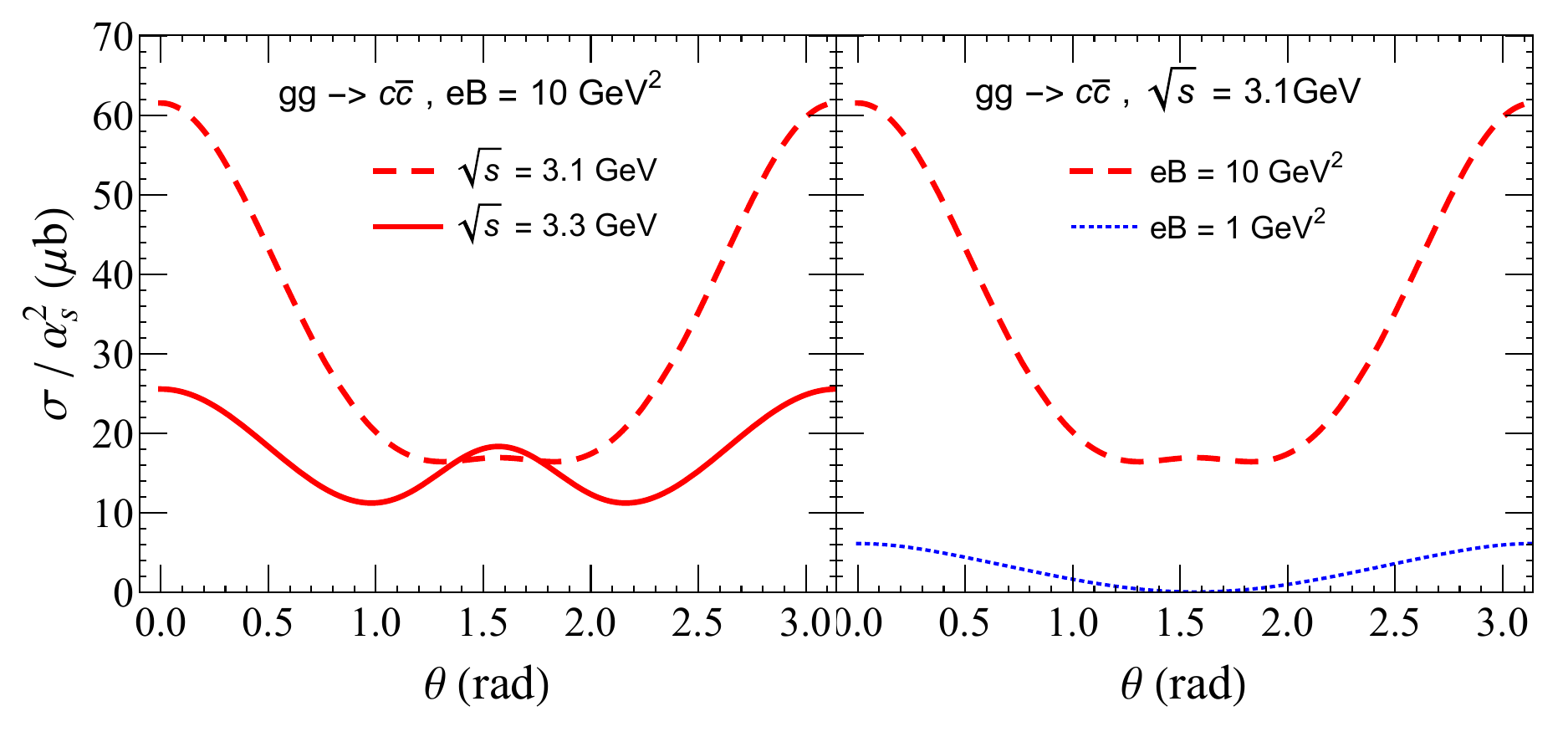}
	\caption{The scaled cross section as a function of the polarization angle $\theta$ at fixed magnetic field $eB=1$ GeV$^2$ (left panel) for invariant mass $\sqrt s =3.3$ (solid line) and $3.1$ (dashed line) GeV and at fixed colliding energy $\sqrt s=3.1$ GeV (right panel) for magnetic field $eB=1$ (dotted line) and $10$ (dashed line) GeV$^2$. }
	\label{fig3}
\end{figure}

The polarization angle dependence of the cross section is shown in Fig.~\ref{fig3}. To guarantee a sizeable cross section, we choose the colliding energy $\sqrt s$ to be inside the narrow region above the threshold. The cross section reaches its maximum at $\theta=0, \pi$. Since the maximum is characterized by the QCD process ($s$ channel), the gluon self-interaction plays the dominant role in heavy quark production in a strong magnetic field. From the left panel at fixed magnetic field, the sizeable cross section at $\sqrt s=3.1$ GeV jumps down rapidly already at $\sqrt s=3.3$ GeV! This indicates a very narrow window for the cross section in a strong magnetic field. On the other hand from the right panel, the production strongly increases with the magnetic field.
%------------------------------
%------------------------------

%%%%%%%%%%%%%%%%%%%%%%%%%%%%
\subsection{Transverse momentum spectrum in nuclear collisions}
%%%%%%%%%%%%%%%%%%%%%%%%%%%%
Qualitatively speaking, from the center-of-mass energy dependence of the elementary cross section, the magnetic field will lead to an enhancement at low $p_T$ and a suppression at high $p_T$ for the heavy quark production. We now quantitatively calculate the magnetic field effect on the heavy quark distributions in high energy nuclear collisions. We use the obtained cross section for the elementary process $gg\to c\bar c$ to compute the yield and spectrum of the heavy quark pairs.

The differential cross section of heavy quark pairs produced in a proton-proton collision can be factorized as
\begin{eqnarray}
{d^3\sigma_{gg\to c\bar c}^{pp} \over dp_T^2dy_cdy_{\bar c}}=x_1x_2f_g(x_1,Q^2)f_g(x_2,Q^2){d\sigma_{gg\to c\bar c} \over d\hat t}(\hat s, \hat t,\hat u),
\label{eq.dis}
\end{eqnarray}
where the Mandelstam variables $\hat s,\ \hat t,\ \hat u$ are now used for the proton-proton collision and
\begin{eqnarray}
\hat s&=&x_1x_2s, \nonumber\\
\hat t&=&-p_T^2 \left(1+e^{y_{\bar c}-y_c}\right), \nonumber\\
\hat u&=&-p_T^2 \left(1+e^{y_c-y_{\bar c}}\right)
\end{eqnarray}
for the elementary process $gg\to c\bar c$, $x_1$ and $x_2$ are momentum fractions carried by the two initial gluons which can be expressed in terms of the heavy quark and antiquark rapidities~\cite{Wang:1991hta},
\begin{eqnarray}
x_1&=&{p_T\over \sqrt s}\left(e^{y_c}+e^{y_{\bar c}}\right), \nonumber\\
x_2&=&{p_T\over \sqrt s}\left(e^{-y_c}+e^{-y_{\bar c}}\right),
\end{eqnarray}
and $Q^2=p_T^2$ is the momentum transfer in the proton-proton collision.

Considering the nuclear geometry, the heavy quark production in the early stage of a nucleus-nucleus collision A+B can be expressed as a superposition of the nucleon-nucleon collisions,
\begin{eqnarray}
{d^3\sigma_{gg\to c\bar c}^{AB} \over dp_T^2dy_cdy_{\bar c}}=\int d^2{\bm x}_Tdz_A dz_B \rho_A( {\bm x}_T+{{\bm b}\over2},z_A)\rho_B( {\bm x}_T-{{\bm b}\over 2},z_B) \mathcal {R}^A_g \mathcal {R}^B_g{d^3\sigma_{gg\to c\bar c}^{pp} \over dp_T^2dy_cdy_{\bar c}},
\end{eqnarray}
where $\rho_A$ and $\rho_B$ are the nucleon distributions in the projectile and target nucleus, which are usually taken as the Woods–Saxon distribution, ${\bm x}_T$ and $z_A (z_B)$ are the transverse and longitudinal coordinates of the two colliding nucleons, and ${\bm b}$ is the impact parameter of the nuclear collision. The nuclear shadowing effect on initial gluons is embedded in the factor $\mathcal R_g^{A(B)}$, which depends on the gluon momentum fraction and position~\cite{Helenius:2012wd}. To simplify the calculation, we neglect in this study all the cold nuclear matter effect and take approximately ${\mathcal R}_g^{A(B)}=1$. In this case, the differential cross section is simplified as
\begin{eqnarray}
{d^3\sigma_{gg\to c\bar c}^{AB} \over dp_T^2dy_cdy_{\bar c}} = T_{AB}({\bm b}){d^3\sigma_{gg\to c\bar c}^{pp} \over dp_T^2dy_cdy_{\bar c}},
\label{eq.spectAA}
\end{eqnarray}
where $T_{AB}({\bm b})=\int T_{A}({\bm x}_T-{{\bm b}/2})T_B({\bm x}_T+{{\bm b}/2})d^2{\bm x}_T$ with the thickness functions $T_A$ and $T_B$ is the number of effective nucleon-nucleon collisions at fixed impact parameter ${\bm b}$.

The differential cross section for the element process can be written as
\begin{eqnarray}
{d\sigma_{gg\to c\bar c} \over d\hat t}={|{\mathcal M}_{gg\to c\bar c} |^2 \over 16\pi \hat s^2}.
\label{eq.dsigmaM}
\end{eqnarray}
In vacuum, we have~\cite{Combridge:1978kx},
\begin{eqnarray}
{|{\mathcal M}_{gg\to c\bar c} |^2 \over \pi^2 \alpha_s^2}&=&{12\over \hat s^2}(m^2-\hat t)(m^2-\hat u)+{8\over 3}\left({m^2-\hat u\over m^2-\hat t}+{m^2-\hat t\over m^2-\hat u} \right)-{6\over \hat s}(2m^2-\hat t-\hat u) \nonumber\\
&&-{16m^2\over 3}\left [{m^2+\hat t\over (m^2-\hat t)^2}+{m^2+\hat u\over (m^2-\hat u)^2} \right] +{6\over \hat s} {m^2(\hat t-\hat u)^2\over (m^2-\hat t)(m^2-\hat u)} \nonumber\\
&&-{2\over 3} {m^2(\hat s-4m^2)\over (m^2-\hat t)(m^2-\hat u)},
\end{eqnarray}
and in a strong magnetic field under the LLL approximation, the scattering amplitudes is expressed as
\begin{eqnarray}
{|\mathcal M_{gg\to c\bar c}|^2 \over \pi^2\alpha_s^2} &=& 24 m^2\cos^2\theta\left[{1\over \hat s}+{\sin^2\theta\left(\hat s\sin^2\theta/4 +(\hat t+\hat u)/2+(\hat t-\hat u)^2/(\hat s\cos^2\theta)\right)\over \left(\hat s\sin^2\theta/2 +\hat t+\hat u\right)^2-\left(\left(\hat t^2-\hat u^2\right)/(\hat s\cos\theta)\right)^2}e^{-\hat s\sin^2\theta\over 8|qB|}\right]\nonumber\\
&&+\frac{64}{3}m^2\sin^4\theta\Bigg[{\left(-\sqrt{\hat s}\cos\theta/2+(\hat t-\hat u)/(\sqrt{\hat s}\cos\theta)\right)^2\over \left(\hat s\sin^2\theta/2+\hat t+\hat u-\left(\hat t^2-\hat u^2\right)/(\hat s\cos\theta)\right)^2}\nonumber\\
&&+{\left(\sqrt{\hat s}\cos\theta/2+(\hat t-\hat u)/(\sqrt{\hat s}\cos\theta)\right)^2\over \left(\hat s\sin^2\theta/2+\hat t+\hat u+\left(\hat t^2-\hat u^2\right)/(\hat s\cos\theta)\right)^2}\nonumber\\
&&-{1\over 4}{(\hat t-\hat u)^2/(\hat s\cos^2\theta)-\hat s\cos^2\theta/4\over \left(\hat s\sin^2\theta/2 +\hat t+\hat u\right)^2-\left(\left(\hat t^2-\hat u^2\right)/(\hat s\cos\theta)\right)^2}\Bigg]  e^{-\hat s\sin^2\theta\over 4|qB|}.
\end{eqnarray}
%---------------------------------------------------------------------
\begin{figure}[!htb]
	$$\includegraphics[width=0.55\textwidth]{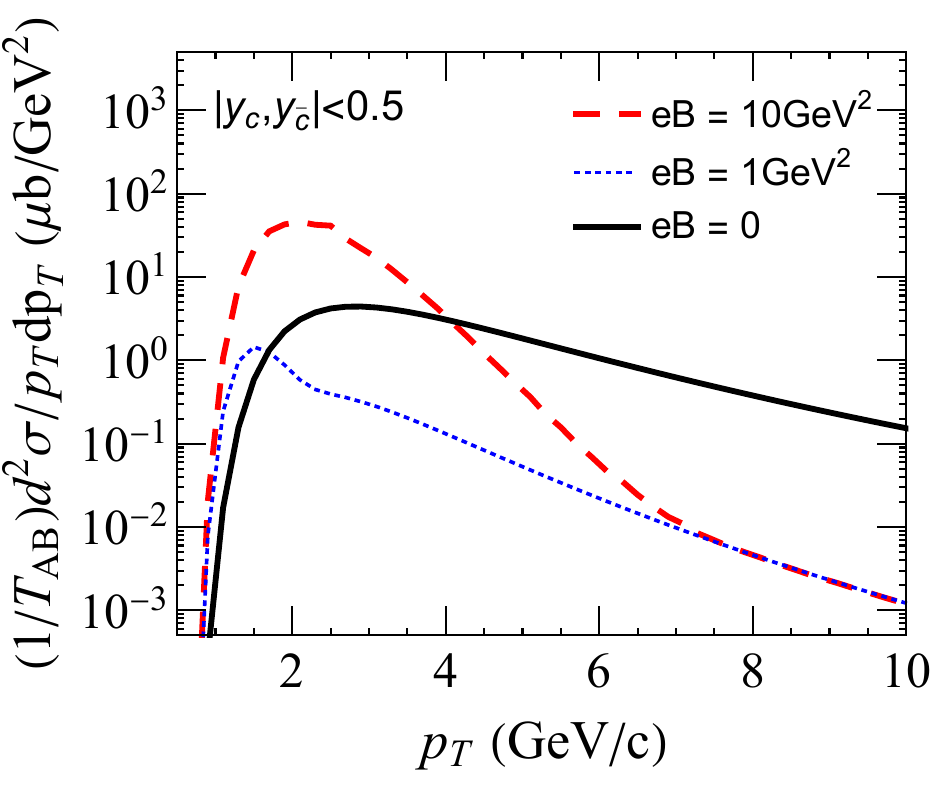}$$
	\caption{The transverse momentum distribution of heavy quark pairs $c\bar c$ in nucleon-nucleon collisions at colliding energy $\sqrt {s_{NN}}= 5.02$ TeV. The heavy quark rapidity is in the region $-0.5<y_c, y_{\bar c} <0.5$, and the magnetic field is fixed to be $eB=0$ (solid line), $1$ (dotted line) and $10$ (dashed line) GeV$^2$.}
	\label{fig4}
\end{figure}
%---------------------------------------------------------------------

Taking the integration over the gluon polarization angle $\theta$, we obtain the $p_T$ spectrum of heavy quark pairs produced in nucleon-nucleon collisions at colliding energy $\sqrt s =5.02$ TeV, shown in Fig.\ref{fig4}. The heavy quark rapidity is chosen to be in the central rapidity $-0.5<y_c, y_{\bar c}<0.5$ and the magnetic field is fixed to be $eB=0, 1$ and $10$ GeV$^2$. Considering the fact that the magnetic field created in nuclear collisions is perpendicular to the beam, the momentum $p_z$ along the magnetic field is just the transverse momentum $p_T$. In the calculation, the gluon distribution $f_g$ is taken from Ref.~\cite{Duke:1983gd}, and the parameters are fitted by the data on deep-inelastic scattering.

In comparison with the result in vacuum with $eB=0$, the heavy quark enhancement at low $p_T$ and suppression at high $p_T$ in a strong magnetic field are resulted from the concentration of heavy quark production in a narrow energy window for the elemental process $gg\to c\bar c$, shown in Fig.\ref{fig2}.

%%%%%%%%%%%%%%%%%%%%%%%%%%%%
\section{Beyond the Lowest Landau Level}
\label{sec4}
\begin{figure}[!htb]
\centering
\includegraphics[width=1.0\textwidth]{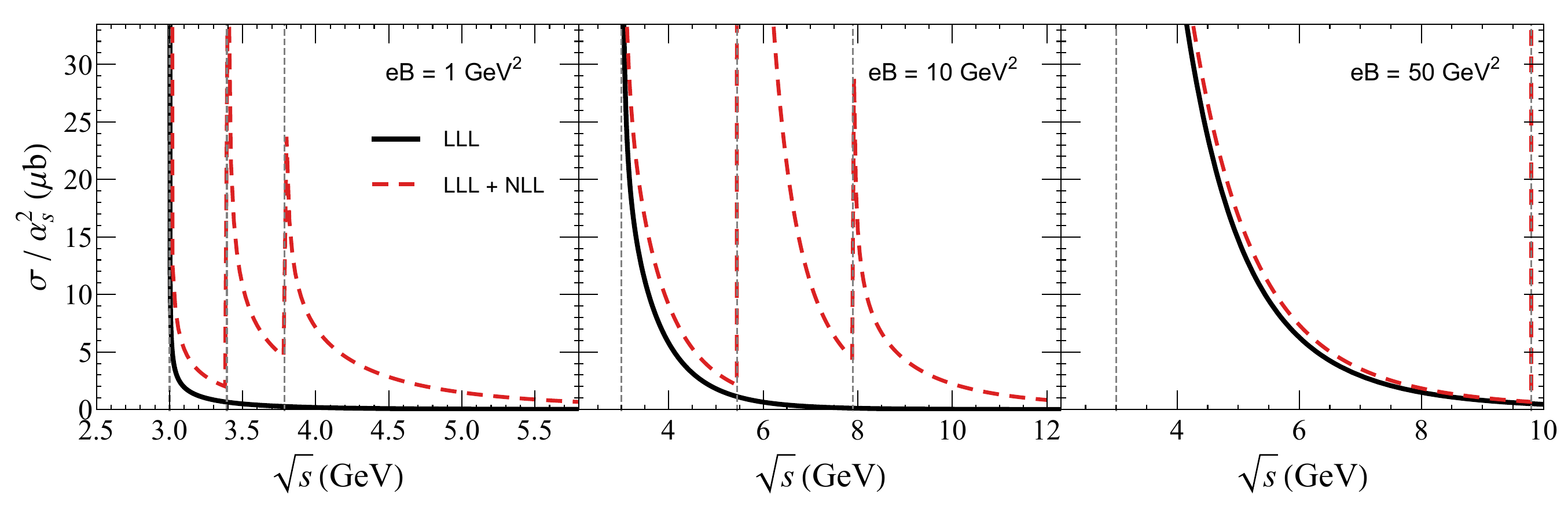}
\caption{The scaled cross section for the elemental process $gg\to c\bar c$ as a function of invariant mass $\sqrt s$ for magnetic field $eB=1$ (left panel), $10$ (middle panel) and $50$ (right panel) GeV$^2$ under the approximation of LLL (solid lines) and NLL (red dashed lines). }
\label{fig5}
\end{figure}

%%%%%%%%%%%%%%%%%%%%%%%%%%%%
In the calculations above we considered a strong magnetic field and took only the Lowest Landau Level with $n=0$ for heavy quarks. A natural question is how good the LLL approximation is. Since an external magnetic field breaks down the translation invariance of a system, and the particle thermal motion restores the invariance, the competition between the magnetic field and the thermal motion will reduce the magnetic effect. Therefore, even if the magnetic field is strong enough, the LLL approximation is not a good approximation at high temperature ~\cite{Huang:2022fgq}. For heavy quark production in high energy nuclear collisions, it happens in the early stage when the system is not yet thermalized. While in this case there is no cancellation from the parton thermal motion, we still need to know the condition for the use of LLL approximation. To this end, we calculate the cross section for the element process $gg\to c\bar c$ to the Next Landau Level (NLL) with $n=0,1$. The result and the comparison with LLL are shown in Fig.\ref{fig5}.

Let's first analyze the new divergences above the threshold $\sqrt s =2m$. The Landau level of the internal quark in the $t-$ and $u-$ channel does not change the divergence structure, and the new divergences come from the Landau levels of the final state heavy quarks. For the calculation with $n=1$, the cross section contains three parts: Both the two final state quarks are with $n=0$ (but the internal quark is with $n=1$), one quark is with $n=1$ and the other with $n=0$, and both the two quarks are with $n=1$. From the collision kinetics, these three parts have different threshold energies,
\begin{eqnarray}
\sqrt{s^{(1)}_{th}} &=& 2m,\nonumber\\
\sqrt{s^{(2)}_{th}} &=& m+\sqrt{m^2+2|qB|},\nonumber\\
\sqrt{s^{(1)}_{th}} &=& 2\sqrt{m^2+2|qB|}.
\end{eqnarray}
The infrared divergence at $p_z=0$ happens at these thresholds. As we discussed above, these divergences at the Landau levels are due to the dimension reduction, they are general phenomena in magnetic field and appear in quarkonium dissociation~\cite{Hu:2022ofv} and other dynamical processes~\cite{Wei:2021dib}.
%------------------------------
%------------------------------

At low energy between the first and second divergences $\sqrt{s_{th}^{(1)}} <\sqrt s < \sqrt{s_{th}^{(2)}}$, the difference between LLL and NLL is not big and the LLL is the dominant contribution, especially when the magnetic field is strong enough $eB\gtrsim 10$ GeV$^2$. At $eB=50$ GeV$^2$, the difference almost disappears in the energy region of $\sqrt s < 10$ GeV, and the contribution from the excited states with $n\geq 1$ can be safely neglected. Considering the fact that the partons in a nucleon or a nucleus distribute mainly in the small $x$ region, the heavy quark production at low $p_T$ in nucleus-nucleus collisions with magnetic field $eB\sim 10$ GeV$^2$ can be reasonably well described at the LLL level.

When the two final state quarks are with different energies, namely when one quark is with $n=0$ and the other with $n=1$, the internal gluon in the $t-$channel and $u-$channel may satisfy the on-shell condition, when the magnetic field is strong enough. The magnetic field to guarantee the off-shell gluon should satisfy the condition $|qB|=2eB/3\le 4m^2$ which leads to the maximum field $eB = 13.5$ GeV$^2$. If the magnetic field exceeds this threshold, the gluon decay into a pair of quarks should be considered, and this will lead to other effects which are not included in this study. Therefore, in the case of $eB=50$ GeV$^2$, the center-of-mass energy is restricted to be $\sqrt s<10$ GeV in Fig.\ref{fig5}.

%%%%%%%%%%%%%%%%%%%%%%%%%%%%
\section{Summary}
\label{sec5}
%%%%%%%%%%%%%%%%%%%%%%%%%%%%
Different from light quarks which can be produced in the late formed fireball in high energy nuclear collisions, heavy quarks $c$ and $b$ are generated only in the initial stage of the collisions when the magnetic field is just created and maintains its strongest strength and the system is also not yet thermalized. Therefore, heavy quarks carry the initial properties of the system and are a sensitive probe of the magnetic field. We calculated in this work the heavy quark production at the leading order under an external magnetic field. The magnetic field enters the dynamical process through the internal quark propagator and the external lines for the final state heavy quarks, both are determined by the solution of the Dirac equation.

From the analytic and numerical calculations, we obtain the following three conclusions on the heavy quark production. 1) The gluon self-interaction dominates the elementary production process $gg\to c\bar c\ (b\bar b)$, which is unique in QCD calculations and disappears in QED. 2) The dimension reduction of fermions in an external magnetic field leads to divergences for the elementary cross section at the discrete Landau energy levels. As a consequence, the heavy quark pair production in nucleus-nucleus collisions will be extremely enhanced at low transverse momentum and suppressed at high transverse momentum. 3) The translation invariance is broken and only the momentum along the magnetic field is conserved. Since the isotropic property of the system is violated, the heavy quark production depends strongly on the moving direction.

The above characteristics of heavy quark production in magnetic field may change the experimentally measured quarkonium production in high energy nuclear collisions, for instance the production mechanism and the collective phenomena. In general there are two sources for the quarkonium production, the initial production via hard process controls the high $p_T$ quarkonia and the regeneration in the quark-gluon-plasma phase dominates the low $p_T$ quarkonia~\cite{Zhao:2020jqu}. When the heavy quark $p_T$ is suppressed by the magnetic field, there will be more low $p_T$ quarkonia via the regeneration process and they will carry more properties of the quark-gluon-plasma. On the other hand, the direction dependence of heavy quark production in magnetic field will change the elliptic flow of the final state quarkonia. The elliptic flow is originally introduced by the geometry in non-central nuclear collisions. When the heavy quarks are not fully thermalized with the medium, their anisotropic production in the initial stage will be reflected in the quarkonium production.\\

\noindent {\bf Acknowledgement}: We would like to thank S. Shi and X. Guo for useful discussions. The work is supported by the NSFC grants No. 12075129, the Guangdong Major Project of Basic and Applied Basic Research No.2020B0301030008, the funding from the European Union’s Horizon 2020 research, and the innovation program under grant agreement No. 824093 (STRONG-2020).
\vspace{1cm}
\bibliography{Ref}
\bibliographystyle{JHEP.bst}

\end{document}